\documentclass[aps,prc,twocolumn,amsmath,superscriptaddress,floatfix,nofootinbib]{revtex4-1}

\usepackage{amsmath,amssymb,amsfonts,mathtools,bm}
\usepackage{graphicx}
\usepackage{xcolor,ulem}
\usepackage{braket}
\usepackage{ragged2e} 

\usepackage[caption=false]{subfig}
\usepackage{hyperref}
\hypersetup{
    bookmarksnumbered=true,bookmarksopen=true,colorlinks=true,
    citecolor=blue,linkcolor=red,urlcolor=blue
}

\begin{document}
\title{Impact of nuclear triaxial deformation on electromagnetic fields in relativistic $^{129}\mathrm{Xe}+^{129}\mathrm{Xe}$ collisions}

\author{Maidi Huang}\email{241120031@fzu.edu.cn}
\affiliation{Department of Physics, Fuzhou University, Fujian 350116, China}

\author{Jin Hu}\email{hu-j23@fzu.edu.cn}
\affiliation{Department of Physics, Fuzhou University, Fujian 350116, China}

\author{Yunpeng Liu}\email{yunpeng.liu@tju.edu.cn}
\affiliation{Department of Physics, Tianjin University, Tianjin 300350, China}

\author{Baoyi Chen}\email{baoyi.chen@tju.edu.cn}
\affiliation{Department of Physics, Tianjin University, Tianjin 300350, China}
\affiliation{
The International Joint Institute of Tianjin University,
Fuzhou, Tianjin University, Tianjin 300072, China}

\begin{abstract}
Electromagnetic fields produced in relativistic heavy-ion collisions depend sensitively on the initial spatial distribution of nuclear charge. Using the Liénard--Wiechert potential with a triaxially deformed Woods--Saxon density, we calculate the transverse electric and magnetic field distributions in $^{129}\mathrm{Xe}+{}^{129}\mathrm{Xe}$ collisions at $\sqrt{s_{NN}}=5.44\text{ TeV}$. We systematically examine how the triaxiality angle $\gamma$ modifies the field structure at the collision time ($t=0$) in semi-central events, using spherical nuclear collisions as a baseline. The results show that nuclear triaxiality causes distinct spatial redistributions of both electric and magnetic fields in the transverse plane. These findings indicate that initial electromagnetic fields encode key information on intrinsic nuclear shapes, offering an additional constraint on nuclear deformation and its consequences for field-sensitive observables in heavy-ion collisions.
\end{abstract}

\maketitle

\section{Introduction}
Relativistic heavy-ion collisions produce the strongest electromagnetic fields created in laboratory environments~\cite{Rafelski:1975rf,Skokov:2009qp,Voronyuk:2011jd,Deng:2012pc}. Because these fields originate directly from fast-moving spectator and participant protons, their magnitude and spatial profile are dictated by the initial charge distribution inside the colliding nuclei~\cite{Cheng:2019qsn}. Mapping the initial electromagnetic fields is therefore essential for two reasons: it offers a unique tool to probe intrinsic nuclear structure—including quadrupole deformation~\cite{Zhu:2026vql,Zhao:2026tgn} and neutron skins~\cite{Li:2018wuv,Zhang:2025voj,Tian:2025wbe,Zhao:2026zno}—and it provides the baseline required to quantitatively interpret field-dependent observables, such as the chiral magnetic effect~\cite{Voloshin:2004vk,Kharzeev:2007jp,Bzdak:2012ia,Deng:2014uja,Feng:2021oub,ALICE:2022ljz} and the electromagnetic response of the quark--gluon plasma~\cite{Bzdak:2009fc,STAR:2009wot,Huang:2015oca,Gardim:2019xjs,Parkkila:2021tqq,Hu:2022ofv,Grayson:2022asf,Fang:2024skm,Fang:2024sym,Zhao:2026mhx}.

Nuclear deformation introduces spatial anisotropies that leave measurable imprints on final-state observables. For example, deformation alters the geometry of the hot deconfined medium, which can be inferred through anisotropic heavy-quarkonium suppression~\cite{Liu:2026sne}. Compared to hadronic or quarkonium observables that evolve through complex medium dynamics, initial electromagnetic fields are generated almost instantaneously during the passage of the nuclei, making them a much cleaner signature of the initial nuclear geometry. While calculations based on the Liénard--Wiechert potential have incorporated quadrupole deformation~\cite{Skokov:2009qp,Deng:2012pc}, event-by-event fluctuations~\cite{Toneev:2012zx}, and spatial charge distributions~\cite{Giacalone:2020awm,Bally:2021qys,STAR:2024wgy}, the specific impact of triaxial deformation on the spatial structure of electromagnetic fields remains largely unexplored.

The $^{129}\text{Xe}$ nucleus provides an ideal testbed for investigating triaxiality. Situated in the transitional region near the $Z=50$ shell closure~\cite{Gupta:2020rwb,Zhao:2024lpc}, $^{129}\text{Xe}$ exhibits both significant quadrupole deformation and non-zero triaxiality~\cite{Helppi:1981doq}. Modern nuclear density functional theory and low-energy experimental data consistently point to a rigidly triaxial ground-state configuration for $^{129}\text{Xe}$~\cite{Bally:2021qys,Bally:2022rhf}. This intrinsic shape is parameterized by the overall deformation magnitude $\beta$~\cite{Giacalone:2019pca} and the triaxiality angle $\gamma$. Varying $\gamma$ redistributes proton density among the intrinsic principal axes of the nucleus without changing its overall volume. In relativistic $^{129}\text{Xe}+{}^{129}\text{Xe}$ collisions, this spatial redistribution of charge can noticeably modify the initial electric and magnetic field profiles~\cite{ATLAS:2022dov}.

In this work, we calculate the initial electromagnetic fields in relativistic $^{129}\mathrm{Xe}+{}^{129}\mathrm{Xe}$ collisions by combining a triaxially deformed Woods--Saxon charge distribution with the Liénard--Wiechert potential. We specifically examine how nuclear deformation modifies the transverse spatial profiles of the electric and magnetic fields. Our results show that varying the triaxiality angle $\gamma$ systematically redistributes both fields, changing their magnitude and transverse symmetry. These calculations quantify how sensitive initial electromagnetic fields are to nuclear triaxiality, providing a useful benchmark for future efforts to identify experimentally accessible observables tied to intrinsic nuclear geometry~\cite{Klein:2020fmr}.

\section{Theoretical model}

\begin{figure}[htp!] 
\centering 
\includegraphics[width=0.47\textwidth]{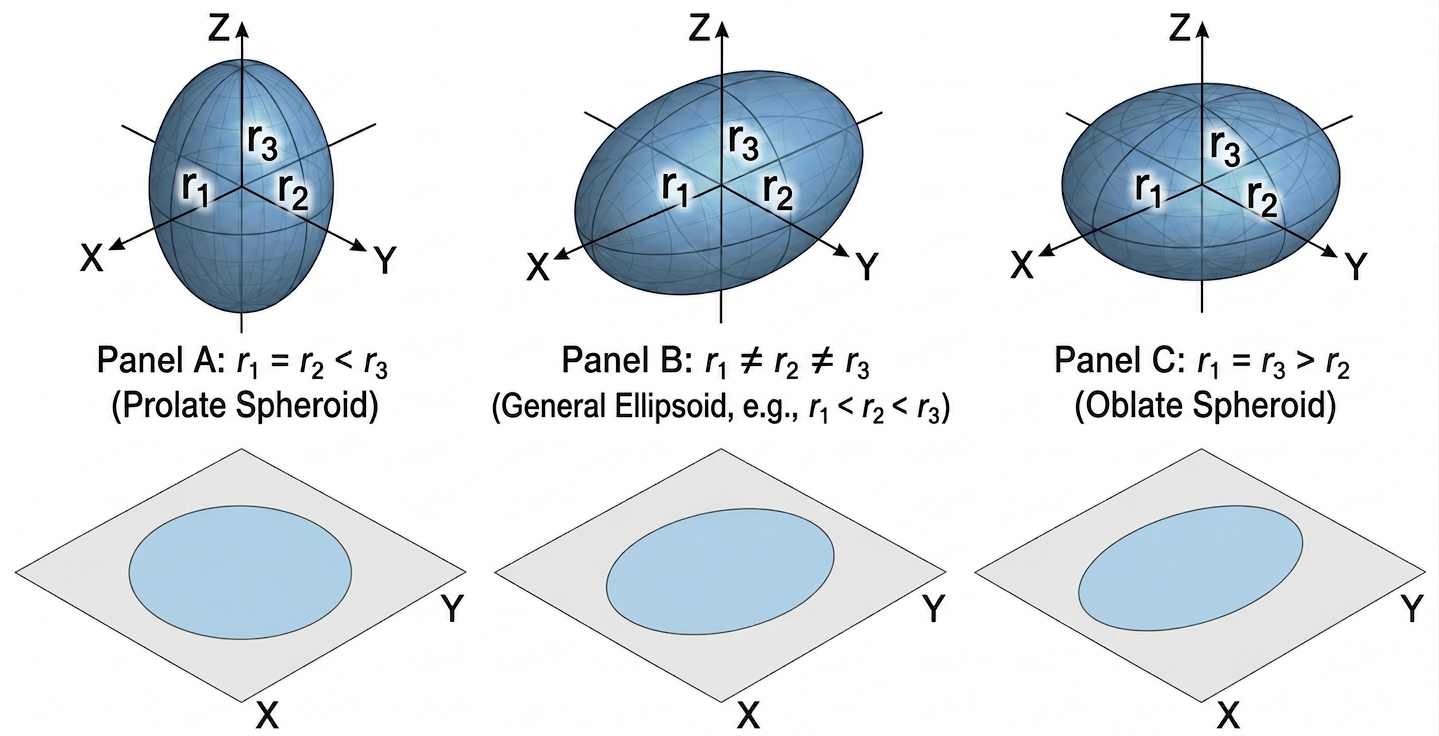}
\caption{(A) A prolate spheroid ($r_1 = r_2 < r_3$), featuring axial symmetry along the $z$-axis and a circular transverse projection.(B) A general triaxial ellipsoid ($r_1 \neq r_2 \neq r_3$), lacking axial symmetry and yielding an orientation-dependent elliptical projection.(C) An oblate spheroid ($r_1 = r_3 > r_2 $), exhibiting axial symmetry perpendicular to the $z$-axis and an elliptical transverse profile.
} 
\label{Fig.1} 
\end{figure}

To calculate the electromagnetic fields produced by a nucleus, the triaxially deformed Woods--Saxon nuclear density~\cite{Woods:1954zz} is defined in the nuclear rest frame $\mathbf r'=(x',y',z')$ as follows~\cite{Miller:2007ri,Shou:2014eya,Ismail:2016zzq,Lu:2023fqd}:

\begin{equation}
\rho(r',\theta',\varphi')
=
\frac{\rho_0}
{1+\exp\left[
\dfrac{r'-R(\theta',\varphi')}{a}
\right]}.
\label{lab-eq-rho1}
\end{equation}
Here, $a$ denotes the nuclear skin diffuseness parameter~\cite{Anni:1994ey,Shou:2014eya}. $\rho_0$ controls the central density of nucleons~\cite{Hirano:2009ah,Loizides:2014vua}. The orientation-dependent radius function, $R(\theta',\varphi')$, is scaled by the half-density radius $R_0$ and incorporates key structural parameters that govern nuclear deformation. For a localized nucleus exhibiting quadrupole deformations, the shape-dependent radius can be expanded in terms of spherical harmonics as follows~\cite{Heinz:2004ir,Verney:2025efj}:
\begin{align}
R(\theta',\varphi')
=
R_0
&[
1+\beta
(
\cos\gamma\,Y_{20}(\theta',\varphi') \nonumber \\
&
+
\frac{\sin\gamma}{\sqrt{2}}
\left[
Y_{22}(\theta',\varphi')
+
Y_{2,-2}(\theta',\varphi')
\right]
)
]
\end{align}
where $Y_{\ell m}(\theta',\varphi')$ denote the standard spherical harmonics. Here, $\beta$ measures the overall quadrupole deformation of the nucleus, while $\gamma$ controls its triaxiality—that is, how much the three principal axes differ from one another \cite{Davydov:1958zz,Shou:2014eya,Bonatsos:2025nbo}. As shown in Fig.~\ref{Fig.1}, $\gamma = 0^\circ$ corresponds to a prolate shape, whereas $\gamma = 60^\circ$ represents an oblate shape. In the intermediate range, $0^\circ < \gamma < 60^\circ$, the nucleus exhibits full triaxiality \cite{STAR:2024wgy}, characterized by three unequal principal axes, see Fig.~\ref{Fig.1}. 

We focus on $^{129}\text{Xe}$, which nuclear density functional theory calculations predict to be rigid triaxial~\cite{Bally:2021qys,Bally:2022rhf,Jia:2021qyu}. Situated in the $A \approx 130$ transitional region near the $Z = 50$ magic shell closure, the xenon isotopic chain exhibits strong shape coexistence and pronounced triaxiality \cite{Garrett:2021kfb,Bonatsos:2024vqp,Yang:2023dql,Guliyev:2022szr}. Here, we investigate how modulating the triaxiality parameter $\gamma$ affects the electromagnetic fields generated in $^{129}\mathrm{Xe}+{}^{129}\mathrm{Xe}$ collisions \cite{Scamps:2020fyu,CMS:2019cyz}.  In our computational framework, the Woods--Saxon parameters are fixed at $a = 0.49\text{ fm}$, $R_0 = 5.6\text{ fm}$, and a central density of $\rho_0 = 0.15\text{ fm}^{-3}$~\cite{Anni:1994ey}. To isolate the effect of quadrupole deformation, we set $\beta = 0$ for the spherical baseline and $\beta = 0.21$ for the deformed configuration~\cite{Bally:2021qys}. The triaxiality angle is varied among $\gamma = 0^\circ$, $30^\circ$, and $60^\circ$, representing prolate, maximally triaxial, and oblate geometries, respectively. Assuming the proton density traces the total nuclear density, we express it as $\rho_p(\mathbf r',\theta',\varphi') = \frac{Z}{A}\rho(\mathbf r',\theta',\varphi')$~\cite{Miller:2007ri}, which directly enters the electromagnetic field calculations.

In the moving nucleus, the electric and magnetic fields from each small volume are calculated using the Liénard--Wiechert potential
~\cite{Deng:2012pc,Skokov:2009qp,Deng:2014uja,McLerran:2013hla},
\begin{align}
e\,d \mathbf E_{i}(t,\mathbf r)
&=
\frac{e^2}{4\pi}
\frac{
\left(1-v^2\right)
\left(
\mathbf R_{i}-R_{i}\mathbf v
\right)
}
{
\left(
R_{i}-\mathbf R_{i}\cdot\mathbf v
\right)^3
}
\,d q_i ,
\label{eq:LW_E}
\\
e\,d\mathbf B_{i}(t,\mathbf r)
&=
\frac{e^2}{4\pi}
\frac{
\left(1-v^2\right)
\mathbf v\times\mathbf R_{i}
}
{
\left(
R_{i}-\mathbf R_{i}\cdot\mathbf v
\right)^3
}
\,d q_i ,
\label{eq:LW_B}
\end{align}
where $d q_i$ represents the electric charge within the $i$-th subvolume $d V_i'$, obtained by dividing the nucleus. ${{\bf R}_i}$ is defined as the relative position from the field point $\bf{r}$ to the source point ${\bf r}_{i}$, and $R_{i}=|{{\bf R}_i}|$ is the scalar length of ${{\bf R}_i}$,where ${\bf r}_i$ is the position of the $i$-th particle with velocity $\bf{v}$ at the retarded time $t_{ret}=t-|{\bf{r}}-{{\bf r}_i}|$\cite{Deng:2012pc,Skokov:2009qp}.

\begin{figure}[htp!]
\includegraphics[width=0.46\textwidth]{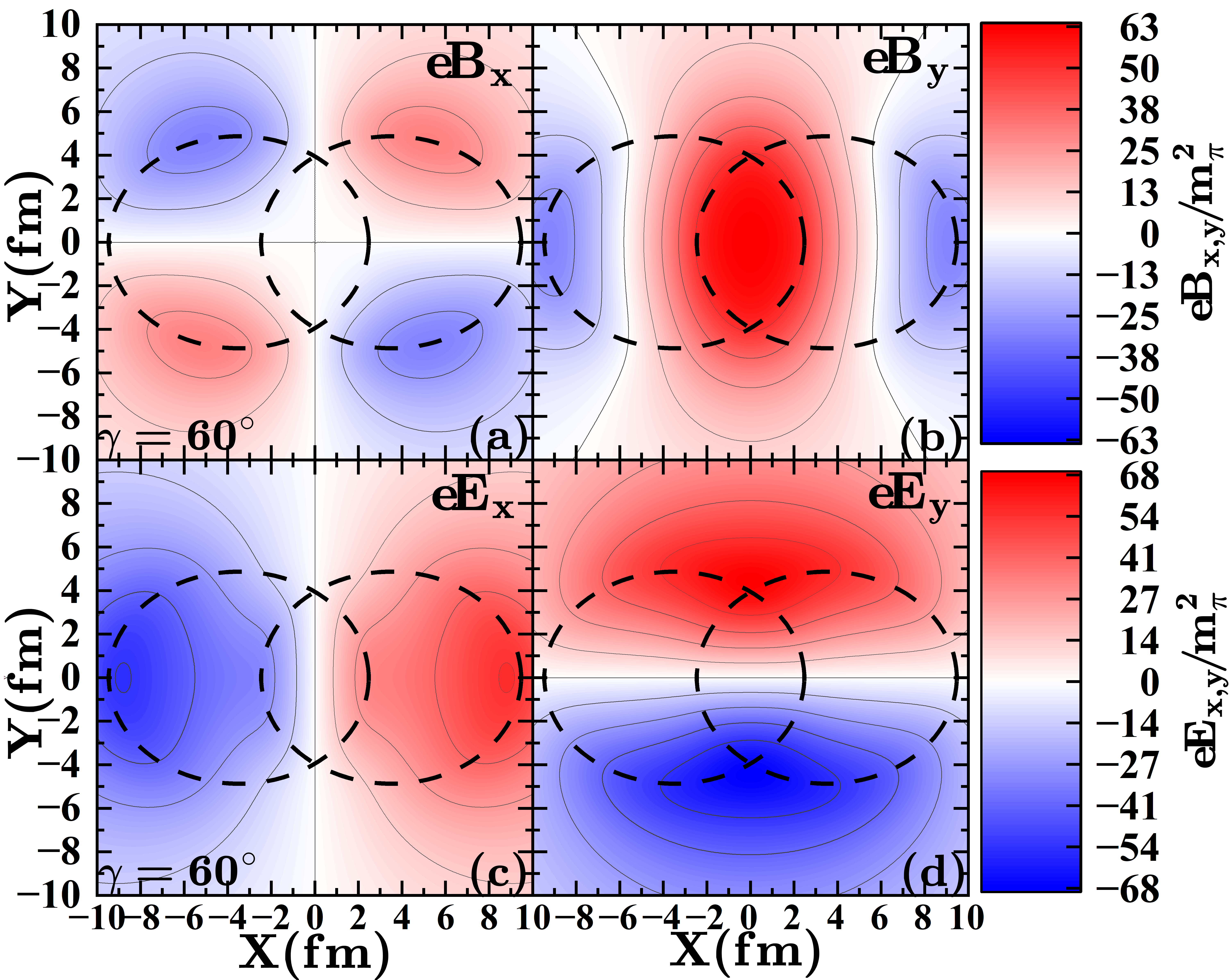}
\caption{ 
Transverse plane of electromagnetic fields scaled with the square of pion mass from numerical calculations generated by 
\(^{129}\mathrm{Xe}+^{129}\mathrm{Xe}\) collisions with the deformed nuclear charge distribution Eq.(\ref{lab-eq-rho1}) at \(t=0~\mathrm{fm}/c\) and impact parameter\(b=7~\mathrm{fm}\), where $\beta = 0.21$, $\gamma = 60^\circ$. Positive values in the figure indicate that the electromagnetic field is directed in the positive direction of the x or y axis, while negative values indicate that the electromagnetic field is directed in the negative direction of the x or y axis. The dashed circles indicate the two  colliding nuclei.
}
\label{Fig.2}
\end{figure}

The total electric and magnetic fields are obtained by summing the contributions from all charge elements of nucleus 1 and nucleus 2, with the definition given as
\begin{align}
\label{eq-two-E}
e\mathbf E(t,\mathbf r)
&=
\sum_{i}
e\,d \mathbf E_{i,1}(t,\mathbf r) + 
\sum_{i} 
e\,d \mathbf E_{i,2}(t,\mathbf r), \\
\label{eq-two-B}
e\mathbf B(t,\mathbf r)
&=
\sum_{i}
e\,d \mathbf B_{i,1}(t,\mathbf r) +
\sum_{i}
e\,d \mathbf B_{i,2}(t,\mathbf r).
\end{align}
Fig.~\ref{Fig.2} shows the transverse plane distribution of the electriec and magnetic field intensities generated by the collision of $^{129}\text{Xe}+^{129}\text{Xe}$, and normalized to the square of the pion mass $m_\pi^2$.
The calculated electric-field components, (\(eE_x\), \(eE_y\)), and
magnetic-field components, (\(eB_x\) ,\(eB_y\)), exhibit the symmetry expected from the Liénard--Wiechert potential. As shown in the upper panel of Fig.~\ref{Fig.2}, $eB_x$ remains small (indicated by the near-white color), whereas $eB_y$ peaks in the collision region (highlighted in red). In contrast, the electric field components ($eE_x$ and $eE_y$) are nearly zero within the collision region but become significant in surrounding areas, as evidenced by the the red and blue regions in the lower panel.

\section{Effects of nuclear deformation on electromagnetic fields}

Based on the above theoretical model, we study the influence of different triaxial angles \(\gamma\) on the transverse distribution of the electromagnetic field at $t=0$ in \(^{129}\)Xe + \(^{129}\)Xe collision with collision energy $\sqrt{s_{NN}}=5.44 \text{ TeV}$ and impact parameter $b=7$ fm. The value of the electromagnetic field is the superposition of the electromagnetic fields generated by the two atomic nuclei at that point. The effect of the nuclear deformation on the magnetic fields is quantified as
\begin{equation}
\label{eq-diff-EB}
e\Delta B=eB{(\gamma,\beta=0.21)}-eB{(\beta=0)},
\end{equation}
where $B$ represents the magnetic fields  obtained from Eq.(\ref{eq-two-B}) and similarly the electric fields $E$ follows from Eq.(\ref{eq-two-E}). The fields are evaluated at three representative triaxiality parameters ($\gamma=0^\circ$, $30^\circ$, and $60^\circ$).

\begin{figure}[!htpb] 
\centering 
\includegraphics[width=0.47\textwidth]{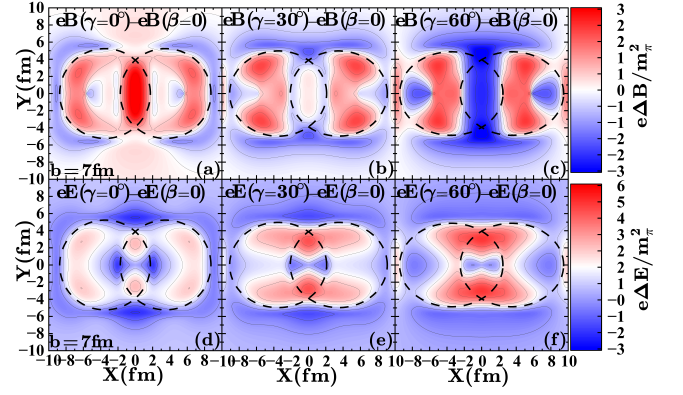} 
\caption{
Transverse spatial profiles of the change in the magnetic field (upper panel) and electric field (lower panel) with different values of triaxiality angle \(\gamma\). It is generated in $\sqrt{s_{NN}}=5.44\text{ TeV}$ \(^{129}\)Xe+\(^{129}\)Xe collisions. The three columns from left to right correspond to evolving triaxiality angle \(\gamma\) parameters of $\gamma = 0^\circ$, $30^\circ$, and $60^\circ$, respectively. 
}
\label{Fig.3} 
\end{figure}

To illustrate the impact of nuclear deformation on electromagnetic fields, Fig.~\ref{Fig.3} displays the field differences computed via Eq.~(\ref{eq-diff-EB}), where all changes (increases or decreases) are evaluated relative to the spherical nucleus case. As $\gamma$ increases, the deviation from the spherical baseline evolves as expected, driven by the redistribution of proton density under triaxial deformation. In the collision region which is the overlap of two nuclei, the magnetic field is enhanced at $\gamma = 0$ but becomes progressively suppressed with increasing $\gamma$. Consequently, the field difference $e\Delta B$ shifts from positive to negative values, as reflected by the color transition from red to blue. In contrast, the electric field exhibits a more complex spatial profile resulting from the coupled deformations of both colliding nuclei. In the overlap area of two nuclei, the spatial distribution of the electric field changes with $\gamma$, suggesting that distinct nuclear deformation modes may influence the dynamical evolution of partons in the deconfined medium.

\begin{figure}[htp!] 
\centering 
\includegraphics[width=0.48\textwidth]{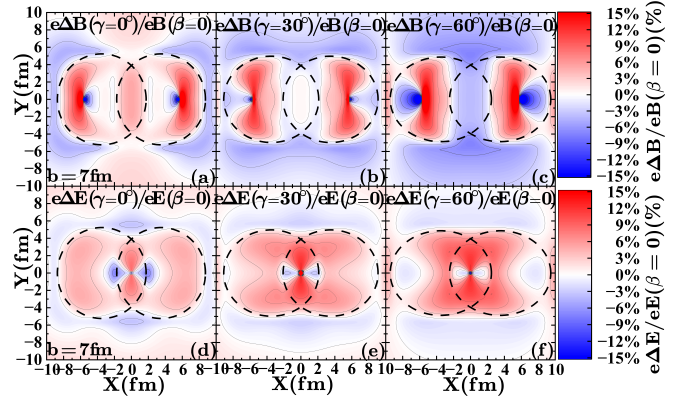}
\caption{Relative changes in the magnetic (upper panel) and electric (lower panel) fields in the transverse plane for different values of the triaxial angle $\gamma$. All changes are evaluated relative to a spherical nucleus.
} 
\label{Fig.4} 
\end{figure}

Fig.~\ref{Fig.4} shows the relative changes in the magnetic and electric fields, which are defined as $\Delta B/B(\beta=0)$ and $\Delta E/E(\beta=0)$, respectively. Here, $B(\beta=0)$ and $E(\beta=0)$ represent the magnetic and electric fields in the spherical case.
In the spectator-nucleon regions, the relative change in the magnetic field can reach about 15\%, as indicated by the red and blue regions in the figure. For the electric field, it increases by $3\%-6\%$ in the collision area for $\gamma=0^\circ$ and by $6\%-12\%$ for $\gamma=60^\circ$, confirming the overall increasing trend. This significant difference in the electromagnetic fields induced by the nuclear deformation may also alter the production of photoproduced particles.

\section{Conclusion}

In summary, we have investigated the effect of nuclear deformation on the initial electromagnetic fields generated in relativistic $^{129}\mathrm{Xe}+{}^{129}\mathrm{Xe}$ collisions. The fields are calculated by applying the Liénard--Wiechert potentials to a triaxially deformed Woods--Saxon proton density distribution. Focusing on fixed nuclear orientations, we systematically evaluate the initial transverse electromagnetic fields and analyze their dependence on the deformation parameters $\beta$ and $\gamma$. Relative to a spherical baseline, nuclear deformation introduces distinct spatial modifications to both fields. The spatial distributions of both the magnetic and electric fields are markedly altered by nuclear deformation, as characterized by varying deformation parameters. This modification, in turn, also influences dilepton photoproduction.

\vspace{0.5cm}
\begin{acknowledgments}
We thank Jiamin Liu for the helpful discussions. 
This work is supported by the National Natural Science Foundation of China under Grant Nos.~12575149 , 12175165 and 12505149.
\end{acknowledgments}

\bibliographystyle{apsrev4-1}
\bibliography{ref.bib}


\end{document}